\documentclass[3p]{elsarticle}
\pdfoutput=1 
\usepackage{graphicx}
\usepackage{amsfonts}
\usepackage{amsmath}
\usepackage{url}
\usepackage{epstopdf}
\usepackage{color}
\usepackage{bm}
\usepackage{multirow}
\usepackage{epsfig}

\newtheorem{definition}{Definition}[section]

\newtheorem{theorem}[definition]{Theorem}

\DeclareMathAlphabet\mathbit
    \encodingdefault\rmdefault\bfdefault\itdefault
\DeclareOldFontCommand{\bi}{\normalfont\bfseries\itshape}{\mathbit}

\newcommand{\be}{\begin{equation}}
\newcommand{\ee}{\end{equation}}

\def\fakebold#1{\relax\ifvmode\leavevmode\fi%
\ifmmode%
\setbox0=\hbox{$#1$}%
\else%
\setbox0=\hbox{#1}%
\fi%
\kern-.02em\copy0 \kern-\wd0%
\kern .04em\copy0 \kern-\wd0%
\kern-.0125em\raise.02em\box0%
}%

\def\ve{\varepsilon}

\def\beq{\begin{eqnarray}}
\def\eeq{\end{eqnarray}}

\def\ve{\varepsilon}
\def\dc{\partial}
\def\bc{\begin{center}}
\def\ec{\end{center}}
\def\bt{\begin{tabular}}
\def\et{\end{tabular}}

\def\R{{\mathbb R}}
\def\ctan{\mathrm{ctan}}
\def\ctanh{\mathrm{ctanh}}

\def\w{w}
\def\x{\bm{x}}
\def\diam{\mathrm{diam}}

\begin{document}

\begin{frontmatter}

\title{Resonance scattering in a waveguide with identical thick barriers}
\author{Andrey Delitsyn}
  \ead{delitsyn@mail.ru}
\address{Kharkevich Institute for Information Transmission Problems of RAN, Moscow, Russia}
\address{National Research University Higher School of Economics, Myasnitskaya Street 20, Moscow 101000, Russia}

\author{Denis S. Grebenkov}
 \ead{denis.grebenkov@polytechnique.edu}
\address{Laboratoire de Physique de la Mati\`ere Condens\'ee, \\
CNRS -- Ecole Polytechnique, IP Paris, 91128 Palaiseau, France}

\address{Institute of Physics \& Astronomy, University of Potsdam, 14476 Potsdam-Golm, Germany}

\begin{abstract}
We consider wave propagation across an infinite waveguide of an
arbitrary bounded cross-section, whose interior is blocked by two
identical thick barriers with holes.  When the holes are small, the
waves over a broad range of frequencies are almost fully reflected.
However, we show the existence of a resonance frequency at which the
wave is almost fully transmitted, even for very small holes.  This
resonance scattering, which is known as tunneling effect in quantum
mechanics, is demonstrated in a constructive way by rather elementary
tools, in contrast to commonly used abstract methods such as searching
for complex-valued poles of the scattering matrix or non-stationary
scattering theory.  In particular, we derived an explicit equation
that determines the resonance frequency.  The employed elementary
tools make the paper accessible to non-experts and educationally
appealing.
\end{abstract}


\end{frontmatter}

\section{Introduction}

Since classic works by Rayleigh \cite{1}, scattering in waveguides is
known to exhibit resonance features.  The resonance character can be
curvature-induced, be related to scattering near the waveguide cut-off
frequency or to a resonance in the waveguide cross-section, or
originate from obstacles forming a resonator that is weakly coupled to
the waveguide
\cite{Parker67,Evans94,Duclos95,Exner96,Bulla97,Davies98,Linton07,Hein08,Grebenkov13}.
In the latter case, the resonator can be either joint to the waveguide
from outside, or made inside.  The former setting is used in mufflers:
incoming waves at frequencies far from the resonance one are almost
fully transmitted; in contrast, the waves near the resonance frequency
are almost fully reflected.  In the second setting, when barriers with
small holes are inserted inside the waveguide, the situation is
different.  If there is a single barrier with Dirichlet boundary
condition, the incoming wave cannot ``squeeze'' through a small hole
and is thus almost fully reflected.  Intuitively, putting more
barriers might seem to help further blocking the wave transmission.
However, if there are two barriers, they can form a resonator, which
is coupled to the waveguide, so that an incoming wave at the resonance
frequency can be almost fully transmitted.  This somewhat
counter-intuitive effect in acoustics is known as tunneling effect in
quantum mechanics \cite{2}.

In spite of a large amount of works on resonance scattering in physics
literature, most of them were focused on approximate computations of
the wave transmission coefficients (see, e.g., \cite{3}).  In turn,
mathematical aspects of resonance scattering of the last type have
been less studied (see \cite{Evans94,Bulla97,Davies98} and references
therein).  One can also mention several works by Arsen'ev \cite{4,5,6}
who applied non-stationary scattering theory.  While the geometric
structure of the problem can be rather general, the derived results
are typically formulated in the form of an alternative: either
scattering is resonant at a given frequency, or this frequency
corresponds to a trapping mode.  Another technique of asymptotic
expansions was applied by Sarafanov and co-workers \cite{7,8} in order
to analyze the limiting behavior in planar waveguides when the size of
a hole in two barriers goes to zero.  It is worth emphasizing that the
mathematical proofs in these works are rather complicated.

In this paper, we provide a much simpler analysis of the resonance
scattering problem for the case of a waveguide of arbitrary constant
bounded cross-section with two identical barriers that are
perpendicular to the waveguide axis.  This problem has two small
parameters: the size of the hole and the difference between the wave
frequency and the resonance frequency.  Our goal is to reveal how
these two parameters should be related to ensure wave transmission.
In particular, we show that the width of barriers can be arbitrary
large that may have interesting applications.  Former studies of
resonance transmission commonly relied on the notion of resonances,
i.e., complex-valued poles of the scattering matrix.  We do not use
this notion that facilitates all the proofs.  In fact, our proofs are
constructive and conceptually simple, even though some formulas are
cumbersome.  Showing a possibility of such mathematically simple
proofs in resonance scattering problems is one of the educational
goals of this paper.

\section{Formulation and main result}

\begin{figure}
\centering
\includegraphics[width=0.55\textwidth]{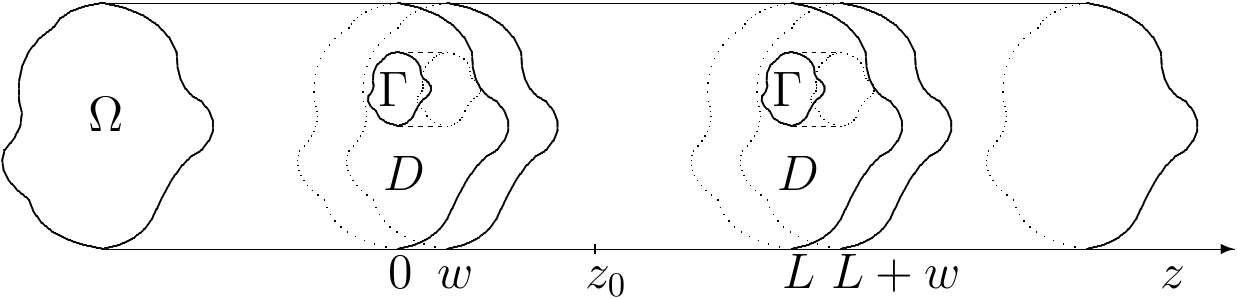}
\caption{
Infinite cylinder $Q = \Omega \times \R$ of a bounded cross-section
$\Omega \subset \R^2$, with two identical thick barriers of
cross-section $D$ and thickness $\w$, separated by distance $L - \w$.
Each barrier has a hole of cross-section $\Gamma = \Omega \backslash
D$.}
\label{fig:domain}
\end{figure}

Let us consider scattering in a waveguide $Q_0$ of a bounded
cross-section $\Omega \subset \R^d$, which contains two identical
barriers of thickness $\w$ separated by distance $L-\w$: $D \times
(0,\w)$ and $D \times (L,L + \w)$, with $D \subset \Omega$
(Fig. \ref{fig:domain}):
\begin{equation}
Q_0 = (\Omega \times \R) \backslash \biggl((D \times (0,\w)) \cup (D \times (L,L+\w)) \biggr) \subset \R^{d+1}.
\end{equation}
We study wave propagation through the waveguide $Q_0$ when the
barriers are closing, i.e., the opening part of the barriers, $\Gamma
= \Omega \backslash D$, is vanishing.  As a similar problem for
infinitely thin barriers was studied in \cite{Delitsyn18}, the main
focus and novelty of the present paper is a finite thickness $\w$ of
barriers.

We consider the stationary wave equation
\be \label{eq:u_eq}
\Delta u + k^2 u = 0 \qquad \textrm{in} ~ Q_0,\ee
with Dirichlet boundary condition on the waveguide walls and on the
barriers,
\be \label{eq:u_BC}
u|_{\dc Q_0} = 0 , \ee
and standard radiation conditions
\begin{subequations}
\begin{align}
u(\x,z) & = e^{i \gamma_1 z} \psi_1(\x) + r_1 e^{-i \gamma_1 z} \psi_1(\x) + \sum \limits_{n=2}^{\infty} r_n e^{\gamma_n z} \psi_n(\x) \qquad (z < 0), \\
u(\x,z) & = t_1 e^{i \gamma_1 z} \psi_1(\x) + \sum \limits_{n=2}^{\infty} t_n  e^{-\gamma_n z} \psi_n(\x) \qquad (z > L + \w),
\end{align}
\end{subequations}
where $r_n$ and $t_n$ are unknown reflection and transmission
coefficients, points in $Q_0$ are written as $(\x,z)$ (with $\x \in
\Omega$ being the transverse coordinate and $z \in \R$ the
longitudinal coordinate along the waveguide axis), $\psi_n(\x)$ and
$\lambda_n$ are the $L_2(\Omega)$-normalized eigenfunctions and
eigenvalues of the Laplace operator in the cross-section $\Omega$:
\be - \Delta \psi_n = \lambda_n \psi_n, \quad \quad \psi_n|_{\dc \Omega} = 0  \quad (n=1,2,3,\ldots), \ee
and
\be \gamma_1 = \sqrt{k^2 - \lambda_1} ,  \quad \gamma_n = \sqrt{\lambda_n - k^2} \qquad (n \geq 2) . \ee
The reflection coefficients can be expressed by using the
orthogonality of eigenfunctions $\{\psi_n\}$:
\be 1 + r_1 = \bigl(u|_{z=0}, \psi_1\bigr)_{L_2(\Omega)}, \qquad r_n = \bigl(u|_{z=0}, \psi_n\bigr)_{L_2(\Omega)}. \ee

In this paper, we prove the following result.
\begin{theorem} 
Let $\delta = \diam\{\Gamma\}$ be the diameter of the opening part
$\Gamma$ of the barriers.  For any fixed wavelength $k$ between
$\sqrt{\lambda_1}$ and $\sqrt{\lambda_2}$, we show that 
\begin{equation}
\lim\limits_{\delta\to 0} r_1 = -1 ,
\end{equation}
i.e., the wave is fully reflecting in the limit of closed barriers.
In turn, for any non-empty $\Gamma$ with $\delta > 0$ small enough,
there exists a resonance wavelength $k_D$ at which
\begin{equation}
r_1 \approx 0 ,
\end{equation}
i.e., the wave is almost fully propagating across two almost closed
barriers.  In other words, for any $\ve > 0$ there exists $\delta' >
0$ such that for any $\Gamma$ with $\diam\{\Gamma\} < \delta'$, there
exists $k_D$ such that $|r_1| < \ve$.
\end{theorem}
Moreover, as our proof is constructive, we will derive an equation,
from which the resonance wavelength $k_D$ can be found.

\section{Derivation}

We consider weak solutions of Eq. (\ref{eq:u_eq}) from $
H^{1,loc}(Q_0) $, i.e., the restriction of the solution to any finite
subdomain $Q'$ of $Q_0$ should belong to $H^1(Q')$.  Moreover, the
series determining the solution should converge in $ L_2(\Omega) $.
Under standard conditions on the boundary $\partial Q$, these
solutions are smooth up to regular parts of the boundary.

\subsection{Reduction to two single-barrier problems}

First, we show that the original problem can be reduced to two
problems in a half cylinder with a single barrier:
\begin{equation}
Q = (\Omega \times (-\infty,z_0)) \backslash (D \times (0,\w)) ,
\end{equation}
where $ z_0 = (\w + L)/2 $.

(i) The first problem involves Dirichlet boundary condition on the
cross-section at $ z = z_0 $:
\begin{subequations}
\begin{align}
\Delta u^D + k^2 u^D & = 0  \quad \textrm{in} ~ Q, \\
\label{eq:u_BC1}
\left. u^D\right|_{\dc Q} & = 0, \\
\label{eq:uD_rad}
u^D(\x,z) & = e^{i \gamma_1 z} \psi_1(\x) + r_1^D e^{-i \gamma_1 z} \psi_1(\x)
+ \sum \limits_{n=2}^{\infty} r_n^D e^{\gamma_n z} \psi_n(\x) \qquad (z < 0), \\
\label{eq:uD_z0}
\left. u^D \right|_{z=z_0} & = 0. 
\end{align}
\end{subequations}

(ii) The second problem involves Neumann boundary condition on the
cross-section at $ z = z_0 $:
\begin{subequations}  \label{eq:Neumann}
\begin{align}
\Delta u^N + k^2 u^N & = 0  \quad \textrm{in} ~ Q,\\\
\label{eq:u_BC2}
\left. u^N\right|_{\dc Q} & = 0, \\
u^N(\x,z) & = e^{i \gamma_1 z} \psi_1(\x) + r_1^N e^{-i \gamma_1 z} \psi_1(\x)
+ \sum \limits_{n=2}^{\infty} r_n^N e^{\gamma_n z} \psi_n(\x) \qquad (z < 0), \\
\label{eq:uN_z0}
\left. \frac{\dc u^N}{\dc z}\right|_{z=z_0} & = 0. 
\end{align}
\end{subequations}

From the solutions of these problems, we can construct the solution of
the original scattering problem in $Q_0$.  Indeed, let us extend the
solution of $u^D$ antisymmetrically and the solution $u^N$
symmetrically onto $Q_0$:
\begin{subequations}
\begin{align}
u^D(\x,z) & = - u^D(\x,2z_0 - z)  \quad (z > z_0), \\
u^N(\x,z) & = u^N(\x,2z_0 - z)  \quad (z > z_0).
\end{align}
\end{subequations}
These extensions are the solutions of the Helmholtz equation
(\ref{eq:u_eq}) in $Q_0$, subject to Dirichlet conditions on $\partial
Q_0$ (i.e., on the cylinder walls and on the barriers) and the
following radiation conditions for $ z > L+\w $
\begin{subequations}
\begin{align}
u^D(\x,z) & = - e^{i 2 \gamma_1 z_0} e^{-i \gamma_1 z} \psi_1(\x) - r_1^D e^{-i 2 \gamma_1 z_0} e^{i \gamma_1 z} \psi_1(\x)
- \sum \limits_{n=2} r_n^D e^{2 \gamma_n z_0} e^{-\gamma_n z} \psi_n(\x), \\
u^N(\x,z) & = e^{i 2 \gamma_1 z_0} e^{-i \gamma_1 z} \psi_1 + r_1^N e^{-i 2 \gamma_1 z_0} e^{i \gamma_1 z} \psi_1(\x)
+ \sum \limits_{n=2} r_n^N e^{2 \gamma_n z_0} e^{-\gamma_n z} \psi_n(\x).
\end{align}
\end{subequations}
The half sum of $ u^D $ and $ u^N $ gives the solution of the
original scattering problem, with
\begin{align}  \label{eq:r1}
r_n & = \frac{r_n^N + r_n^D}{2} \qquad (n \geq 1), \\
t_1 & = \frac{r_1^N - r_1^D}{2} e^{-2i\gamma_1z_0} \,,  \qquad  t_n = \frac{r_n^N - r_n^D}{2} e^{2\gamma_n z_0}  \qquad (n \geq 2).
\end{align}

\subsection{Dirichlet problem solution $u^D$}

To find the solution $u^D$ of the first problem in $Q$, let us also
introduce the $L_2(\Gamma)$-normalized eigenfunctions and eigenvalues
of the Laplace operator in the cross-section of the hole, $\Gamma =
\Omega \backslash D$:
\be - \Delta \chi_n = \mu_n \chi_n, \quad \quad \chi_n|_{\dc \Gamma} = 0  \qquad (n=1,2,3,\ldots) \ee
and set
\begin{equation}
\beta_n = \sqrt{\mu_n - k^2}  \quad (n \geq 1).
\end{equation}
The eigenfunctions $\chi_n$ form an orthonormal basis in
$L_2(\Gamma)$.  In the following, we consider that
\be  \lambda_1 < k^2 < \min\{\mu_1,\lambda_2\}  \ee
so that the coefficients $\gamma_n$ and $\beta_n$ are real for all $n
\geq 1$.  Moreover, if the hole $\Gamma$ is small, $\mu_1$ is large so
that $k$ lies between $\sqrt{\lambda_1}$ and $\sqrt{\lambda_2}$.

We can consider a general solution of the Helmholtz equation in the
domain $\Gamma \times (0,\w)$
\be \label{eq:uD_1}
u^D(\x,z) = \sum\limits_{n=1}^\infty \bigl(e_{1n} \sinh (\beta_n (z - \w)) + e_{2n} \sinh (\beta_n z)\bigr) \chi_n(\x), \ee
where the cofficients $A_{1n}$ and $A_{2n}$ can be expressed as
\be e_{1n} = - \frac{(u_0, \chi_n)_{L_2(\Gamma)}}{\sinh(\beta_n \w)}, \qquad 
e_{2n} = \frac{(u_1, \chi_n)_{L_2(\Gamma)}}{\sinh(\beta_n \w)}, \ee
where
\begin{equation}
u_0 = u^D|_{z=0}, \qquad u_1 = u^D|_{z=\w} .
\end{equation}

Similarly, in the domain $\Omega \times (\w,z_0)$, we have
\be \label{eq:uD_2}
u^D(\x,z) = e_1 \sin (\gamma_1 (z - z_0)) \psi_1(\x) + \sum\limits_{n=2}^\infty e_n \sinh(\gamma_n (z - z_0)) \psi_n(\x), \ee
where the coefficients $e_n$ can be expressed as
\be e_1 = - \frac{(u_1, \psi_1)_{L_2(\Gamma)}}{\sin(\gamma_1 \ell)} , \qquad 
e_n = - \frac{(u_1, \psi_n)_{L_2(\Gamma)}}{\sinh(\gamma_n \ell)}  , \ee
where
\begin{equation}
\ell = z_0 - \w = \frac{L - \w}{2} \,.
\end{equation}

{\bf Remark.}
At this moment, we do not discuss the convergence of the series.
Moreover, we will differentiate formally the series without studying
the validity of this operation up to the introduction of Eqs.
(\ref{eq:auxil3}).  These formal steps are just needed as a background
to establish these equations.  The solution of these equations will
solve the problem (\ref{eq:u_BC}).  In fact, if Eqs. (\ref{eq:auxil3})
have a solution in the functional space $ W $ (introduced below in
Eq. (\ref{eq:Wi})), it can be extended to the whole waveguide $ Q $
with the aid of Eqs. (\ref{eq:uD_rad}, \ref{eq:uD_1}, \ref{eq:uD_2}).
Indeed, these series determine $u^D$ in the whole domain $Q$ as an
element of $ H^{1, loc}(Q) $ that satisfies the Helmholtz equation,
boundary and radiation conditions.  The series determining the
radiation condition converges in $L_2$ at any cross-section because $
u \in H^{1, loc}(Q) $.

Using formal representations
\begin{subequations}
\begin{align}
\left. \frac{\dc u^D}{\dc z}\right|_{z=0-0} & = -i \gamma_1 (u_0, \psi_1) \psi_1
+ \sum \limits_{n=2}^\infty \gamma_n (u_0, \psi_n) \psi_n + 2 i \gamma_1 \psi_1, \\
\left. \frac{\dc u^D}{\dc z}\right|_{z=0+0} & = -\sum \limits_{n=1}^\infty \beta_n \biggl(\ctanh(\beta_n \w) (u_0, \chi_n)
- \frac{1}{\sinh(\beta_n \w)} (u_1, \chi_n)\biggr) \chi_n  ,\\
\left. \frac{\dc u^D}{\dc z}\right|_{z=\w-0} & = - \sum \limits_{n=1}^\infty \beta_n
\biggl(\frac{1}{\sinh(\beta_n \w)} (u_0, \chi_n) - \ctanh(\beta_n \w) (u_1, \chi_n) \biggr) \chi_n  ,\\
\left. \frac{\dc u^D}{\dc z}\right|_{z=\w+0} & = - \gamma_1 \ctan(\gamma_1 \ell) (u_1, \psi_1) \psi_1
- \sum \limits_{n=2}^\infty \gamma_n \ctanh(\gamma_n \ell) (u_1, \psi_n) \psi_n,
\end{align}
\end{subequations}
and imposing the continuity of $ \frac{\dc u}{\dc z} $ at $z = 0$ and
$z = \w$, we obtain two functional equations on $u_0$ and $u_1$:
\begin{subequations}  \label{eq:auxil3}
\begin{align} \label{eq:auxil3a}
 -i \gamma_1 (u_0, \psi_1) \psi_1 + A_0 u_0 + C u_1 & = - 2 i \gamma_1 \psi_1, \\
\label{eq:auxil3b}
B u_0 + \gamma_1 \ctan(\gamma_1 \ell) (u_1, \psi_1) \psi_1 + A_1 u_1 & = 0, 
\end{align}
\end{subequations}
where the operators $A_0$, $A_1$, $B$ and $C$ are defined as
\begin{subequations}  \label{eq:ABC_def}
\begin{align} \label{eq:A1_def}
A_0 f & = \sum \limits_{n=2}^\infty \gamma_n \bigl(f, \psi_n\bigr)_{L_2(\Gamma)}  \psi_n 
+ \sum \limits_{n=1}^\infty \hat{\beta}_n  \bigl(f, \chi_n\bigr)_{L_2(\Gamma)}  \chi_n, \\
\label{eq:A2_def}
A_1 f & = \sum \limits_{n=2}^\infty \gamma_n \ctanh(\gamma_n \ell)  \bigl(f, \psi_n\bigr)_{L_2(\Gamma)} \psi_n
+ \sum \limits_{n=1}^\infty \hat{\beta}_n  \bigl(f, \chi_n\bigr)_{L_2(\Gamma)} \chi_n, \\
B f & = - \sum \limits_{n=1}^\infty \frac{\beta_n}{\sinh(\beta_n \w)} \bigl(f, \chi_n\bigr)_{L_2(\Gamma)} \chi_n, \\
C f & = - \sum \limits_{n=1}^\infty \frac{\beta_n}{\sinh(\beta_n \w)} \bigl(f, \chi_n\bigr)_{L_2(\Gamma)} \chi_n, 
\end{align}
\end{subequations}
and
\begin{equation}
\hat{\beta}_n = \beta_n \ctanh(\beta_n \w).
\end{equation}

We understand Eqs. (\ref{eq:auxil3}) and the operators $ A_0, A_1, B,
C $ as follows.
Let us consider two Hilbert spaces
\be  \label{eq:Wi}
W_i = \left\{v \in L_2(\Gamma)~:~  \sum \limits_{n=2}^{\infty} \gamma_n^{(i)} \bigl(v, \psi_n\bigr)_{L_2(\Gamma)}^2 
+ \sum \limits_{n=1}^{\infty} \hat{\beta}_n  \bigl(v, \chi_n\bigr)_{L_2(\Gamma)}^2 < \infty \right\} ,
\quad i = 0, 1,
\ee
with the inner products
\be
(f, g)_{W_i} = \sum \limits_{n=2}^{\infty} \gamma_n^{(i)} \bigl(f, \psi_n\bigr)_{L_2(\Gamma)}\, \bigl(g, \psi_n\bigr)_{L_2(\Gamma)} 
+ \sum \limits_{n=1}^{\infty} \hat{\beta}_n  \bigl(f, \chi_n\bigr)_{L_2(\Gamma)}\, \bigl(g, \psi_n\bigr)_{L_2(\Gamma)},
\ee
where $\gamma_n^{(0)} = \gamma_n$ and $\gamma_n^{(1)} = \gamma_n
\ctanh(\gamma_n \ell)$.  Since $\ctanh(\gamma_n \ell)$ rapidly tend to
$1$ as $n\to\infty$, these functional spaces are equivalent, i.e., any
function belonging to $W_0$, also belongs to $W_1$, and vice-versa.
For this reason, we do not distinguish $W_0$ and $W_1$ and denote
either of them as $W$.  It is easy to see that these functional spaces
are also equivalent to $ H^{\frac{1}{2}}(\Gamma)$ but this equivalence
is not needed in the following.  Note also that for any $v
\in W$,
\begin{equation}  \label{eq:boundW}
||v||_{W} \geq C\, ||v||_{L_2(\Gamma)} ,
\end{equation}
with a strictly positive constant $C$.

The operators $ B $ and $ C $ are bounded in $ L_2(\Gamma)$ and their
norms are small if the diameter of the opening $\Gamma$ is small, see
Sec. \ref{sec:estimates}.
The operators $ A_0 $ and $ A_1 $ in Eqs. (\ref{eq:A1_def},
\ref{eq:A2_def}) can also be rigorously defined; however, for our
purposes, it is sufficient to understand these operators in terms of
the associated quadratic forms, i.e. by setting
\be \bigl(A_i f, g\bigr)_{L_2(\Gamma)} = \bigl(f,g\bigr)_{W}, \quad i = 0,1. \ee
As the operators $A_0$ and $A_1$ are positive definite (in the sense
of positive definite quadratic forms determined by $ A_0$ and $A_1 $),
their inverses $A_0^{-1}$ and $A_1^{-1}$ are well defined (see
discussion in Sec. \ref{sec:estimates}).

Applying $A_0^{-1}$ and $A_1^{-1}$ to Eq. (\ref{eq:auxil3a}) and
Eq. (\ref{eq:auxil3b}) respectively, we rewrite them in a matrix
operator form
\be \underbrace{\left( \begin{array}{cc} I & A_0^{-1} C \cr A_1^{-1} B & I \end{array} \right)}_{ = M}
\left( \begin{array}{c} u_0 \cr u_1 \end{array} \right) +
\left( \begin{array}{c} - i \gamma_1 (u_0, \psi_1) A_0^{-1} \psi_1 \cr \gamma_1
\ctan(\gamma_1 \ell) (u_1, \psi_1) A_1^{-1} \psi_1 \end{array} \right) =
\left( \begin{array}{c} - 2 i \gamma_1 A_0^{-1} \psi_1 \cr 0 \end{array} \right), \label{eq2}  \ee
where $I$ is the identity operator.  We multiply Eq. (\ref{eq2}) by
the operator inverse to the operator $M$, 
\begin{equation*}
M^{-1} = \left( \begin{array}{cc} R_1 & 0 \cr  0 & R_2 \end{array} \right)
\left( \begin{array}{cc} I & - A_0^{-1} C \cr
- A_1^{-1} B & I \end{array} \right),
\end{equation*}
where
\begin{equation}  \label{eq:R_def}
R_1 = (I - A_0^{-1} C A_1^{-1} B)^{-1}, \qquad R_2 = (I - A_1^{-1} B A_0^{-1} C)^{-1}.
\end{equation}
We obtain the functional equations
\begin{subequations}
\begin{align}
u_0 - i \gamma_1 (u_0, \psi_1)  R_1 A_0^{-1} \psi_1 - \gamma_1 \ctan(\gamma_1 \ell) (u_1, \psi_1) R_1 A_0^{-1} C A_1^{-1} \psi_1 & =
- 2 i \gamma_1 R_1 A_0^{-1} \psi_1 , \\
u_1 + i \gamma_1 (u_0, \psi_1) R_2 A_1^{-1} B A_0^{-1} \psi_1 + \gamma_1 \ctan(\gamma_1 \ell) (u_1, \psi_1) R_2 A_1^{-1} \psi_1 & =
2 i \gamma_1 R_2 A_1^{-1} B A_0^{-1} \psi_1 . 
\end{align}
\end{subequations}
Multiplying each of these equations by $\psi_1$ and integrating over
the hole $\Gamma$, we obtain two linear equations which can be written
in a matrix form as
\be  \label{eq:auxil55}
\left( \begin{array}{cc} 1 + a & b \, \ctan(\gamma_1 \ell) \cr  c & 1 + d \, \ctan(\gamma_1 \ell) \end{array} \right)
\left( \begin{array}{c} (u_0,\psi_1) \cr (u_1,\psi_1) \end{array} \right)
= 2 \left( \begin{array}{c} a \cr c \end{array} \right),
\ee
where
\begin{subequations}
\begin{align}  \label{eq:a_def}
a & = - i \gamma_1 (R_1 A_0^{-1} \psi_1, \psi_1)_{L_2(\Gamma)}, \\
b & = - \gamma_1 (R_1 A_0^{-1} C A_1^{-1} \psi_1, \psi_1)_{L_2(\Gamma)} ,\\
c & =  i \gamma_1 (R_2 A_1^{-1} B A_0^{-1} \psi_1, \psi_1)_{L_2(\Gamma)} ,\\
d & = \gamma_1 (R_2 A_1^{-1} \psi_1, \psi_1)_{L_2(\Gamma)} ,
\end{align}
\end{subequations}
Since $(u_i, \psi_1)_{L_2(\Omega)} = (u_i, \psi_1)_{L_2(\Gamma)}$ for
both $i=0,1$ given that $(u_0)|_{D} = (u_1)|_{D} = 0$ according the
boundary condition (\ref{eq:u_BC}), we did not specify the functional
space for these two scalar products.
Inverting the $2\times 2$ matrix in Eq. (\ref{eq:auxil55}), one finds
$(u_0,\psi_1)$ and $(u_1,\psi_1)$.

Taking the limit $z\to 0$ in the radiation condition
(\ref{eq:uD_rad}), multiplying it by $\psi_1$ and integrating over
$\Omega$, the reflection coefficient $r_1^D$ can be expressed as
\be  \label{eq:r1D}
r_1^D = (u_0,\psi_1)_{L_2(\Omega)} - 1 =  \frac{a-1 + (ad-bc-d) \ctan(\gamma_1 \ell)}{a+1 + (ad-bc+d) \ctan(\gamma_1 \ell)} \,.
\ee
This is the main technical result of this paper that determines
resonance scattering properties.

\subsection{Resonance transmission}

It is important to emphasize that the reflection coefficient $r_1^D$
in Eq. (\ref{eq:r1D}) depends on $\ctan(\gamma_1 \ell)$ and on the
coefficients $a,b,c,d$.  Here, $\ctan(\gamma_1 \ell)$ is determined by
the wavelength $k$, the resonator half-length $\ell$, and the shape of
the cross-section $\Omega$, but does not depend on the hole $\Gamma$.
In turn, the coefficients $a,b,c,d$ depend on the hole diameter
$\delta$.  As shown in Sec. \ref{sec:estimates} below, the
coefficients $a,b,c,d$ vanish as the diameter $\delta$ of the hole
$\Gamma$ goes to $0$.  As a consequence, for a fixed wavelength $k$,
we obtain in the limit of the vanishing hole:
\begin{equation}
r_1^D \to -1  \quad (\delta \to 0).
\end{equation}
Repeating the same analysis for the Neumann problem (\ref{eq:Neumann})
(which is fairly similar and thus not provided here), one can show
that $ (u^N|_{z=0} \psi_1)_{L_2(\Omega)} $ is close to zero and thus
\begin{equation}  \label{eq:r1N}
r_1^N = \bigl(u^N|_{z=0},\psi_1\bigr)_{L_2(\Omega)} - 1 \to -1  \quad (\delta \to 0).
\end{equation}
Substituting these expressions into Eq. (\ref{eq:r1}), we get
\begin{equation}
r_1 \to -1   \quad (\delta \to 0),
\end{equation}
i.e., the wave is fully reflected in the limit of two closed barriers,
as intuitively expected.

Let us now consider the case of two {\it almost} closed barriers,
i.e., $\delta$ is small but strictly positive.  By continuity
arguments, one can argue that $r_1$ remains close to $-1$ for most
wavelengths, except for the resonance one.  Indeed, for a fixed hole
$\Gamma$, Eq. (\ref{eq:r1D}) implies 
\be \label{eq:r1D_1}
r_1^D = 1 \ee
under the condition on the wavelength $k$:
\be 1 + d \, \ctan(\gamma_1 \ell) = 0 . \ee
The wavelength $k_D$ determined by this equation, is the resonance
wavelength of the resonator with Dirichlet condition (\ref{eq:uD_z0}),
and it cannot be the resonance frequency of the same resonator with
Neumann condition (\ref{eq:uN_z0}).  As a consequence,
Eq. (\ref{eq:r1N}) is still applicable at $k_D$, and
Eqs. (\ref{eq:r1N}, \ref{eq:r1D_1}) imply
\begin{equation}
r_1 \approx 0 ,
\end{equation}
i.e., the wave almost fully propagates across two almost closed
barriers.  In other words, we have shown that for the hole diameter
$\delta$ small enough, the waveguide is almost totally reflecting for
most wavelengths, except for the resonance wavelength $k_D$, at which
it is almost fully propagating.

\section{Estimates for operators}
\label{sec:estimates}

\subsection{Estimates for operators $A_0^{-1}$ and $A_1^{-1}$}

As the operators $A_0$ and $A_1$ are positive definite (in the sense
of positive definite quadratic forms determined by $ A_0$ and $A_1 $),
their inverses $A_0^{-1}$ and $A_1^{-1}$ are well defined and for all
$f \in L_2(\Gamma) $:
\be ||A^{-1}_i f||_{L_2(\Gamma)} \leq C ||f||_{L_2(\Gamma)}  \qquad (i=0,1), \ee
for some $C > 0$, and
\be ||A^{-1}_i f||_{W} \leq C' ||f||_{L_2(\Gamma)}   \qquad (i=0,1), \ee
for some $C' > 0$.

Indeed, the inverse $ A_0^{-1} f $ is defined in a weak sense as the
solution of the equation
\begin{equation*}
\sum \limits_{n=2}^\infty \gamma_n \bigl(A_0^{-1} f, \psi_n\bigr)_{L_2(\Gamma)} \, \bigl(v, \psi_n\bigr)_{L_2(\Gamma)} + 
\sum \limits_{n=1}^\infty \hat{\beta}_n  \bigl(A_0^{-1} f, \chi_n\bigr)_{L_2(\Gamma)} \, 
\bigl(v, \chi_n\bigr)_{L_2(\Gamma)} = (f, v)_{L_2(\Gamma)}  
\end{equation*}
for any $v \in W_0$.  Substituting $ v = A_0^{-1} f $ into this
equation, we obtain
\begin{align*}
||A_0^{-1} f||^2_{W} &= 
\sum \limits_{n=2}^\infty \gamma_n \bigl(A_0^{-1} f, \psi_n\bigr)_{L_2(\Gamma)} \, \bigl(A_0^{-1} f, \psi_n \bigr)_{L_2(\Gamma)} + 
\sum \limits_{n=1}^\infty \hat{\beta}_n  \bigl(A_0^{-1} f, \chi_n\bigr)_{L_2(\Gamma)} \, 
\bigl(A_0^{-1} f, \chi_n\bigr)_{L_2(\Gamma)} \\
& = \bigl(f, A_0^{-1} f\bigr)_{L_2(\Gamma)} ,
\end{align*}
from which
\begin{equation*}
||A_0^{-1} f||^2_{W} \leq ||f||_{L_2(\Gamma)} \, ||A_0^{-1} f||_{L_2(\Gamma)} \leq C'_0 \, ||f||_{L_2(\Gamma)}  \, ||A_0^{-1} f||_{W} ,
\end{equation*}
where we used (\ref{eq:boundW}).  We conclude that
\begin{equation*}
 ||A_0^{-1} f||_{W} \leq C'_0 \, ||f||_{L_2(\Gamma)}  .
\end{equation*}
A similar bound can be obtained for $ A_1^{-1} $.

\subsection{Estimates for operators $B$ and $C$}

The estimates for operators $B$ and $C$ are much stronger.
Indeed,
\begin{equation*}
||B f||_{L_2(\Gamma)}^2 = \left \|\sum \limits_{n=1}^{\infty} \frac{\beta_n}{\sinh \beta_n} \bigl(f, \chi_n\bigr)_{L_2(\Gamma)} 
\chi_n \right\|_{L_2(\Gamma)}^2  
\leq  2 \sum \limits_{n=1}^{\infty} \frac{\beta^2_n}{\sinh^2 \beta_n} \bigl(f, \chi_n\bigr)_{L_2(\Gamma)}^2
 \underbrace{||\chi_n||_{L_2(\Gamma)}^2}_{=1}  \leq \frac{2\beta^2_1}{\sinh^2 \beta_1}\, ||f||_{L_2(\Gamma)}^2  , 
\end{equation*}
given that $\beta_n$ monotonously grow with $n$, whereas the function
$z/\sinh(z)$ is monotonously decreasing.  We get thus
\begin{equation*}
||B||_{L_2} \leq C \frac{\beta_1}{\sinh \beta_1}  \longrightarrow 0      \qquad \textrm{as} ~ \delta \to 0  .
\end{equation*}
In fact, as the diameter $\delta = \diam\{\Gamma\}$ of the hole
$\Gamma$ vanishes, the eigenvalue $\mu_1$ goes to infinity, implying
very fast decay of $||B||_{L_2}$.  The same analysis holds for $
||C||_{L_2} $.

From the above estimates we deduce that 
\begin{equation*}
||A_0^{-1} C A_1^{-1} B f||_W \to 0  \qquad \textrm{as} ~ \delta \to 0 
\end{equation*}
so that the operator $ R_1 $ defined in Eq. (\ref{eq:R_def}), is
bounded in $W$.  The same is true for $ R_2 $.  

From these estimates we finally obtain that 
\begin{equation}
\bigl(R_1 A_0^{-1} \psi_1, \psi_1\bigr)_{L_2(\Gamma)} \leq C ||\psi_1||_{L_2(\Gamma)}^2 \to 0 \qquad \textrm{as} ~ \delta \to 0 ,
\end{equation}
implying that $a$ given by Eq. (\ref{eq:a_def}), also vanishes as
$\delta\to 0$.  Similar estimates take place for $b$, $c$, and $d$.

\end{document}